\documentclass[onecolumn,superscriptaddress]{iopart}
\usepackage{ulem}
\usepackage[usenames]{color}
\definecolor{simon}{cmyk}{1.0, 0.0, 1.0, 0.0} 

\definecolor{norm}{cmyk}{0,0,0,1}

\definecolor{luca}{cmyk}{0,1,0,0}

\definecolor{mazurenko}{cmyk}{1,0,0,0}

\usepackage{graphicx}
\usepackage{times}
\usepackage{dcolumn}

\newcommand{\BIO}{Ba$_2$IrO$_4$}
\newcommand{\SrO}{Sr$_2$IrO$_4$}

\begin{document}

\title{The electronic structure of the high-symmetry perovskite iridate~\BIO}

\author{S Moser$^{1,2}$, L Moreschini$^2$, A Ebrahimi$^1$, B Dalla Piazza$^1$, M Isobe$^3$, H. Okabe$^3$, J. Akimitsu$^4$,
VV Mazurenko$^5$, KS Kim$^2$, A Bostwick$^2$, E Rotenberg$^2$, J Chang$^1$, HM R{\o}nnow$^1$ and M Grioni$^1$} 

 \address{$^1$Institute of Condensed Matter Physics (ICMP), Ecole Polytechnique F\'ed\'erale de Lausanne (EPFL), CH-1015 Lausanne, Switzerland}

\address{$^2$Advanced Light Source (ALS), Lawrence Berkeley National Laboratory, Berkeley, California 94720, USA}

\address{$^3$National Institute for Materials Science (NIMS), 1-1 Namiki, Tsukuba, Ibaraki 305-0044, Japan}

\address{$^4$Department of Physics and Mathematics, Aoyama Gakuin University, 5-10-1 Fuchinobe, Chuo-ku, Sagamihara, Kanagawa 252-5258, Japan}

\address{$^5$Theoretical Physics and Applied Mathematics Department, Ural Federal University, Mira Str.19,  620002
Ekaterinburg, Russia}

\date{\today}

\begin{abstract}
We report angle-resolved photoemission (ARPES) measurements, density functional and model tight-binding calculations on Ba$_2$IrO$_4$ (Ba-214), an antiferromagnetic ($T_N=230$~K) insulator. Ba-214 does not exhibit the rotational distortion of the IrO$_6$ octahedra that is present in its sister compound Sr$_2$IrO$_4$ (Sr-214), and is therefore an attractive reference material to study the electronic structure of layered iridates. We find that the band structures of Ba-214 and Sr-214 are qualitatively similar, hinting at the predominant role of the spin-orbit interaction in these materials. Temperature-dependent ARPES data show that the energy gap persists well above $T_N$, and favour a Mott over a Slater scenario for this compound.

\end{abstract}

\maketitle

\section{Introduction}

The iridates are a new family of strongly correlated materials, with fascinating physical properties \cite{Cao1998,Kini2006,Moon2008,Kim2008,Kim2009,Ge2011,Qi2011,Wang2011,Fujiyama2012,Haskel2012,Cetin2012,Comin2012}. Unlike $3d$ transition metal oxides (TMOs), dominated by the Coulomb interaction, or $4d$ TMOs, where Hund's rule coupling plays a major role \cite{Georges2013}, the electronic structure of the $5d$ iridates reflects the coexistence of similar Coulomb, crystal-field (CEF) and spin-orbit (SO) coupling energy scales.
As a result, Mott physics and local magnetic moments can emerge in the iridates for values of the Coulomb interaction that are one order of magnitude smaller than in the $3d$ series.

The layered perovskite \SrO~(Sr-214) has attracted considerable attention because of intriguing similarities with the cuprate parent compound La$_2$CuO$_4$ (LCO). Structurally, it exhibits weakly coupled IrO$_2$ square-lattice planes built from corner-sharing IrO$_4$ plaquettes, analogous to the characteristic CuO$_4$ building blocks of the cuprates \cite{Kim2008,Kim2009,Shimura1995,CRAWFORD1994,Kim2012}. The electronic structure is shaped by strong SO coupling, which splits the Ir $5d~t^5_{2g}$ manifold, so that the highest occupied state is a narrow, half-filled $j_{eff}=1/2$ band. 
The Ir spins order into an antiferromagnetic (AFM) state below $T_N=230-240$~K. According to the leading scenario, Sr-214 is an insulator because a Mott gap $\Delta \sim0.06$~eV opens within this band, but the actual origin of the gap is still being debated.
An alternative Slater picture, coupling the onset of magnetic order with the opening of a gap \cite{Gebhard:Principles,arita_ab_2012}, has been advocated by susceptibility, time-resolved optical conductivity and scanning tunneling spectroscopy data \cite{Moon2009,Okabe2011,Hsieh2012,Li2013}.

One may speculate that Sr-214 could be turned into a superconductor by doping, similarly to LCO. However, superconductivity is hindered by weak in-plane ferromagnetism, attributed to the Dzyaloshinsky-Moriya interaction, which arises from a rotational distortion of the IrO$_6$ octahedra (Fig. 1(a)).
Recently, the sister compound \BIO~(Ba-214), with similar physical properties (Table I), was synthesized using high pressure methods \cite{Okabe2011} (Fig. 2(b)). Due to the larger Ba radius, Ba-214 does not exhibit the rotational distortion, and is therefore a more promising parent compound for possible iridate superconductors. Ba-214 also offers the possibility of studying the electronic structure of an undistorted IrO$_2$ square lattice.
\begin{table}
\caption{\label{tab: Table1} Comparison of structural, magnetic and electronic properties in Ba$_2$IrO$_4$ and Sr$_2$IrO$_4$.
$d_b$ and $d_{ap}$ are the in-plane and apical Ir-O distances.}
\begin{tabular}{c|c|c|c|c|c|c|c}
\br
  &$d_{b}$ ({\AA})&$d_{ap}$ ({\AA})&$\theta_{tilt}$&$T_N$ (K)&$\mu$ ($\mu_B$)&$E_a$ (meV)&Ref.\\
\br  
  Ba$_2$IrO$_4$&2.01& 2.15& 0$^{\circ}$ &230&0.34&70& {\bf} \cite{Okabe2011}\\
  Sr$_2$IrO$_4$&1.98&2.06&11$^{\circ}$&230-240&0.33&60& {\bf} \cite{CRAWFORD1994,Shimura1995,Kini2006}\\
\br
\end{tabular}
\end{table}

\begin{figure}
\centering
\includegraphics[width=86mm]{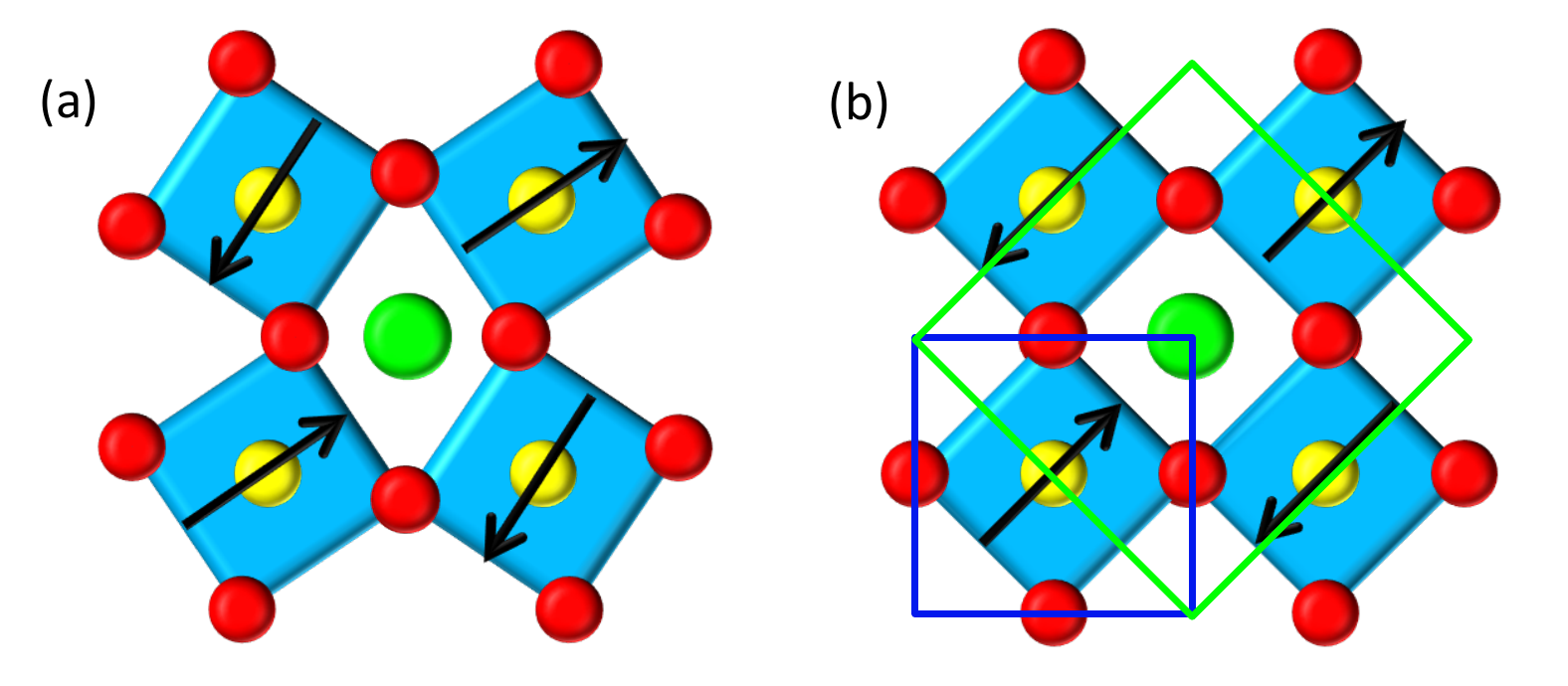}
\caption{\label{fig: Figure1} (a) Sr$_2$IrO$_4$ structure, projected on the $ab$ plane (red=O; green=Sr; yellow=Ir). Apical oxygen atoms are not shown. Each IrO$_6$ octahedron is rotated by 11$^\circ$ around the perpendicular $c$ axis with respect to the ideal K$_2$NiF$_4$ structure, yielding a larger c(2$\times$2) unit cell. The distortion is absent in \BIO~(b) (green=Ba). The arrows illustrate the Ir spin arrangement in the AFM phase. Blue and green squares are the primitive and magnetic unit cell
\cite{Boseggia,Boseggia2}.}
\end{figure}

We present here an investigation of Ba-214~by angle-resolved photoelectron spectroscopy (ARPES). We analyse the experimental data with the help of first-principles density functional (DFT) and model tight-binding (TB) band structure calculations.
We find that the band structure of Ba-214~is quite similar to that of Sr-214, and is therefore rather insensitive to the presence of the rotational distortion.
We also observe a backfolding of the band structure corresponding to a larger $c(2\times2)$ in-plane unit cell, that coincides with the AFM unit cell.
ARPES data collected over a broad temperature range do not give evidence for a temperature-dependent gap, and therefore are more consistent with a Mott than with a Slater scenario.

\section{Methods}
Samples of Ba-214~were grown as in Ref.~\cite{Okabe2011} in the form of dense, black polycrystalline pellets.
The pellets were broken into clusters of $\sim1$mm$ ^3$ size, and then dipped in 1\% hydrofluoric acid for one minute. After rinsing in deionized water, single crystals of $\sim$400 $\mu$m lateral size could be extracted. Crystals naturally exposing the $(001)$ surface were selected under an optical microscope and mounted on ceramic pins.

The ARPES measurements were performed at the electronic structure factory end station of beam line 7 of the Advanced Light Source, Lawrence Berkeley National Laboratory. The combined energy resolution of the monochromator and of the Scienta R4000 hemispherical analyzer was $\sim30$~meV.
Samples were cleaved at $T\sim 100$~K at a pressure $<10^{-10}$~mbar. The Fermi level reference was measured on polycrystalline copper in good electrical equilibrium with the sample. Sample charging hindered measurements below $\sim 80$~K. All data presented here were collected at $T\geq120$~K, where the effect was smaller and under control. The data were subsequently corrected for the residual energy shift, which was estimated from a comparison with data measured in a low-filling mode of the storage ring, with a photon intensity reduced by almost two orders of magnitude.

\begin{figure*}
\centering
\includegraphics[width=135mm]{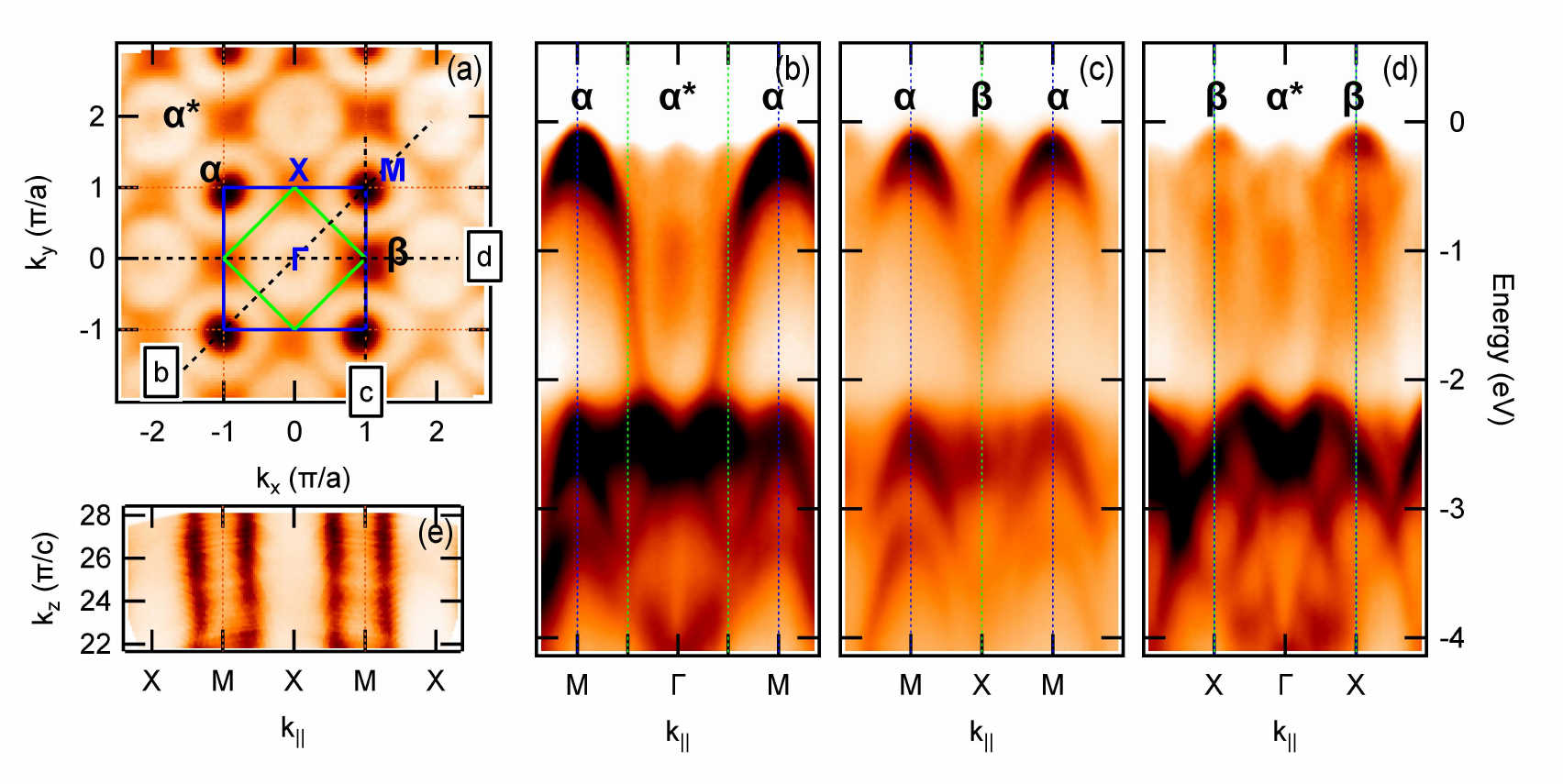}
\caption{\label{fig: Figure2} (a) ARPES $k_x$ vs. $k_y$ constant energy (CE) map of Ba-214, measured at $E=-0.1$~eV with $h\nu=155$~eV at $T=130$~K. It shows intense features centered at the $M$ and $X$ points of the surface BZ (blue square), and a weaker replica of the former at $\Gamma$. The green square is the $c(2\times2)$ BZ.  (b-d) $E$ vs. $k_{\parallel}$ cuts along the high symmetry directions indicated as (b), (c) and (d) in panel~(a) Blue and green vertical lines indicate the boundaries of the $(1\times1)$ and of the $c(2\times2)$ BZs. (e) ARPES $k_x$ vs. $k_z$ CE map at $E=-0.4$~eV and $k_y=\pi/a$, extracted from a photon energy scan between $95$~eV and $162$~eV, assuming an inner potential $V_0=10$~eV. It shows negligible dispersion along the $c$-axis. $\Gamma X=\pi/a=0.78~$\AA$^{-1}$; $\Gamma Z=\pi/c=0.24~$\AA$^{-1}$. In all panels, the darkest features correspond to the largest intensity.}
\end{figure*}

\section{ARPES results}
\subsection{Band structure}

Figure 2(a) presents an ARPES constant energy (CE) map of Ba-214, measured at $E=-0.1$~eV, near the top of the valence band. The map is extracted from a dataset measured at photon energy $h\nu=155$~eV, and $T=130$~K. The blue square is the surface Brillouin zone (BZ) corresponding to the crystallographic unit cell of Fig.\ref{fig: Figure1}(b)\cite{Okabe2011}. The map shows intense round features ($\alpha$ features in the following) at the $M$ points, the corners of the BZ. A second set of features ($\beta$ features), with fourfold symmetry, is observed at the $X$ points. Both $\alpha$ and $\beta$ features are repeated in all BZs of the map. A closer inspection reveals also a weaker, round contour ($\alpha^*$) at all $\Gamma$ points. It will be clear in the following that $\alpha^*$ is a signature of band folding into the smaller $c(2\times2)$ BZ (green square).

Panels 2(b-d) show the experimental $E$ vs. $k_{\parallel}$ dispersion along high-symmetry lines marked (b), (c) and (d) in panel 2(a).
Along $M \Gamma M$, panel 2(b) shows a prominent band with a maximum at the $M$ point, where it gives rise to the $\alpha$ contour. As discussed below, this band corresponds primarily to Ir states of $j_{eff}=3/2$ character. It merges around $-2$~eV with a manifold of O $2p$-derived states. The same band is seen to disperse downwards along the $MXM$ direction in panel 2(c). A second band, with a maximum at the $X$ point, generates the $\beta$ contour. In the Mott scenario, it is assigned to the Ir-derived lower-Hubbard band of $j_{eff}=1/2$ character. The maxima of this band are more visible along the $X \Gamma X$ direction in panel 2(d), which also shows a dispersive feature with a maximum at
$\Gamma$, associated with the $\alpha^*$ contour.
Panel 2(e) shows a $k_x$ vs. $k_z$ CE map for $E=-0.4$~eV and $k_y=\pi/a$. It is extracted from ARPES measurements with photon energies in the range $h\nu=95-162$~eV, assuming an inner potential $V_0=10$~eV. Apart from slight intensity variations with photon energy, the data are essentially independent of $k_z$. Namely, the absence of
wiggling contours indicates that the $k_z$ dispersion at the top of the valence band (VB), and the interplane coupling for these states, are quite small.

\begin{figure}
\centering
\includegraphics[width=50mm]{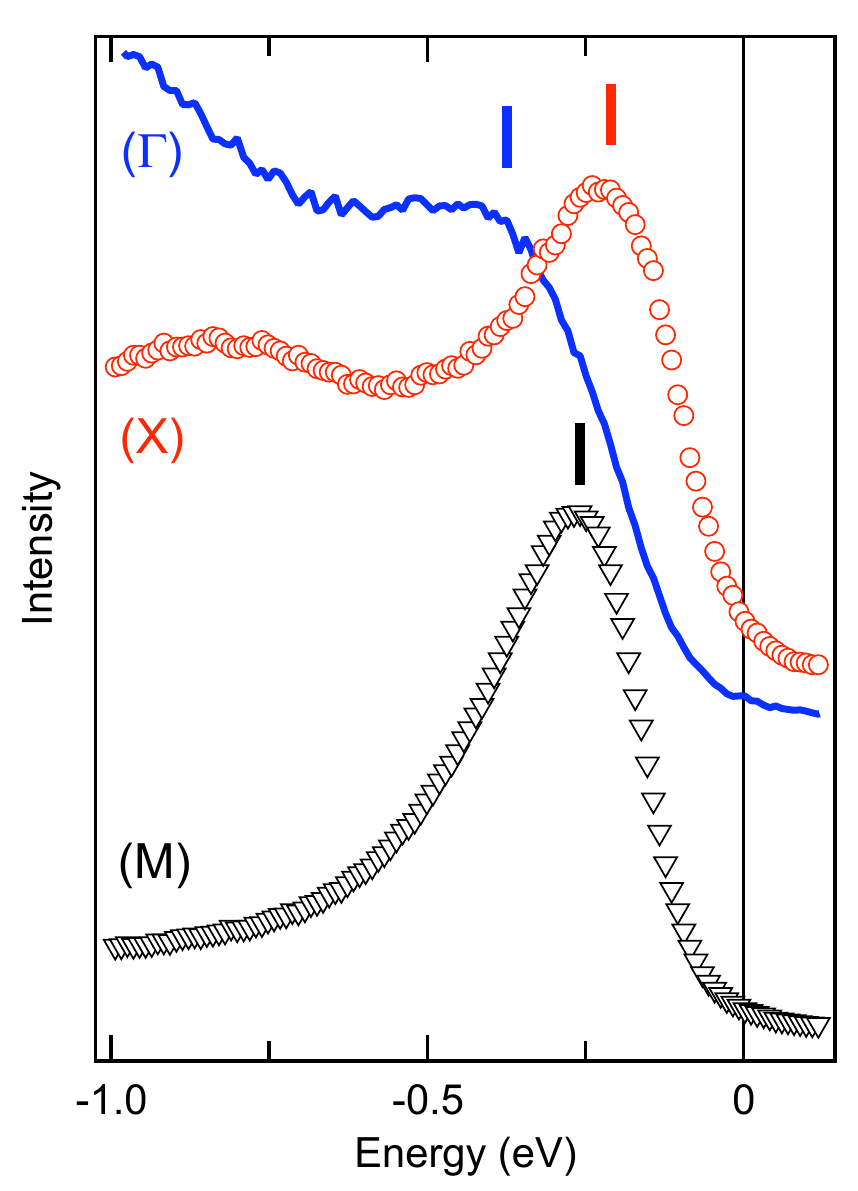}
\caption{\label{fig: Figure3} ARPES spectra of Ba-214, measured at three high symmetry points of the BZ. The corresponding peak positions are indicated by vertical lines.}
\end{figure}

Representative spectra for the $\Gamma$, $M$ and $X$ points of the BZ are shown in Fig. 3. They exhibit rather broad peaks, with maxima at $-0.37$~eV (at $\Gamma$), $-0.26$~eV (at $M$), and $-0.21$~eV (at $X$), which places the VB maximum at the $X$ point. The peak energy at the VB maximum should yield a lower limit for the energy gap, the actual value depending on the separation between the Fermi level and the conduction band mimimum, which cannot be accessed by ARPES. However, the peak binding energy at $X$ ($0.21$~eV) is already larger than the gap value $\Delta_g\simeq 2E_a\sim140$~meV, estimated from the activation energy $E_a=70$~meV of the electrical resistivity \cite{Okabe2011}. This discrepancy, and the broad line shapes, suggest unresolved overlapping features in the spectra of Fig. 3. This hypothesis is supported by the first-principles calculations of Section 4.1. It also explains the different peak energies measured at the top of the band at the $M$ point, and at its backfolded replica at $\Gamma$, since the underlying components can be differently modulated by matrix elements.

\begin{figure}
\begin{center}
\includegraphics[width=100mm]{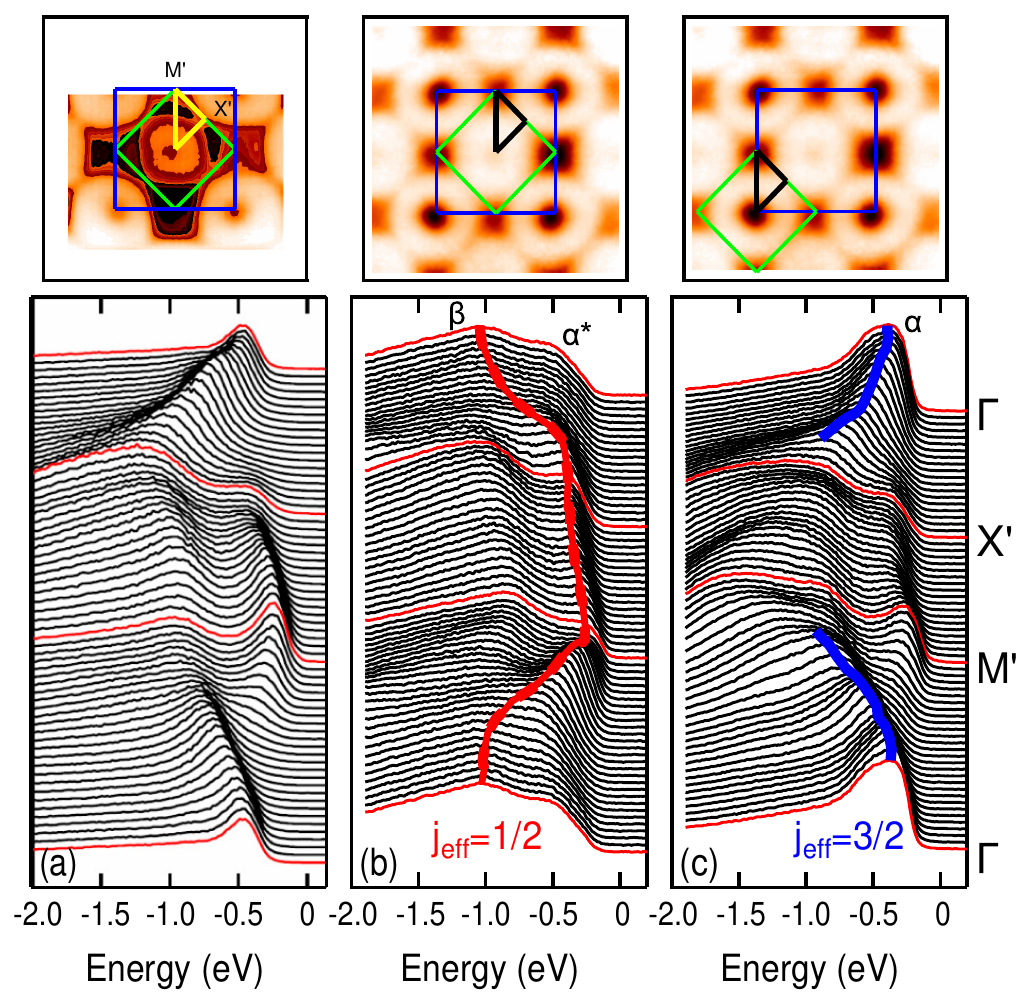}
\caption{\label{fig: Figure4} (color online) (a) ARPES spectra of Sr-214, measured along the $\Gamma M' X'$ contour (yellow triangle) in the $c(2\times2)$ BZ (green square), from Ref.~\cite{Kim2008}. Data for Ba-214, extracted from Fig.~2, are shown along the same contour in (b), and for an equivalent contour in an adjacent $c(2\times2)$ BZ in (c). Spectra corresponding to the high-symmetry points are in red. Matrix elements enhance the signal from $j_{eff}=1/2$ states in (b), and from $j_{eff}=3/2$ states in (c). Thick red (b) and blue (c) curves outline their dispersion.}
\end{center}
\end{figure}

We now compare the electronic structure of Ba-214 and Sr-214. Figure 4(a) reproduces ARPES data on Sr-214 from Ref.~{\cite{Kim2008}. The spectra are measured along the $\Gamma M' X'$ contour in the $c(2\times2)$ BZ, indicated by a yellow triangle in the top panel. Our results for Ba-214, extracted from the dataset of Fig. 2, are shown for the same triangular contour in Fig. 4(b). Data measured along the same triangular contour in the adjacent $c(2\times2)$ BZ are also shown in Fig. 4(c).
Two conclusions can be drawn from the figure. Firstly, there is a good overall correspondence between the band structure of the two compounds.  Secondly, while the relative intensities of the Ba-214 $j_{eff}=3/2$ and $j_{eff}=1/2$ bands in panels 4(b) and 4(c) are different, their dispersions are identical. The triangular contours in 4(b) and 4(c) are equivalent for the
$c(2\times2)$ BZ, but clearly not for the structural BZ (blue square). This shows that -- similarly to the case of Sr-214 \cite{Kim2008} -- the band structure of Ba-214 is folded into the smaller BZ. The smaller intensity of the $\alpha^*$ manifold in 4(b), compared with that of the $\alpha$ manifold in 4(c), is consistent with band folding from a superlattice potential that is substantially weaker than the primary lattice potential \cite{Voit}. The different intensities in the two contours can be exploited to disentangle the two bands. We find that in Ba-214 the width of the $j_{eff}=1/2$ band ($\sim 0.8$~eV) is somewhat larger than in Sr-214 ($\sim 0.5$~eV) \cite{Kim2008}, and is considerably smaller than the width of the $j_{eff}=3/2$ band ($\sim 2.5$~eV). The maxima of the $j_{eff}=1/2$ band at $X$ ($-0.21$~eV) and of the $j_{eff}=3/2$ band at $M$ ($-0.26$~eV) in Ba-214 are shallower than those ($-0.25$~eV and, respectively $-0.45$~eV) of the corresponding bands in Sr-214. Their energy separation is also smaller ($0.05$~eV vs. $0.2$~eV) in Ba-214.

\begin{figure}
\centering
\includegraphics[width=86mm]{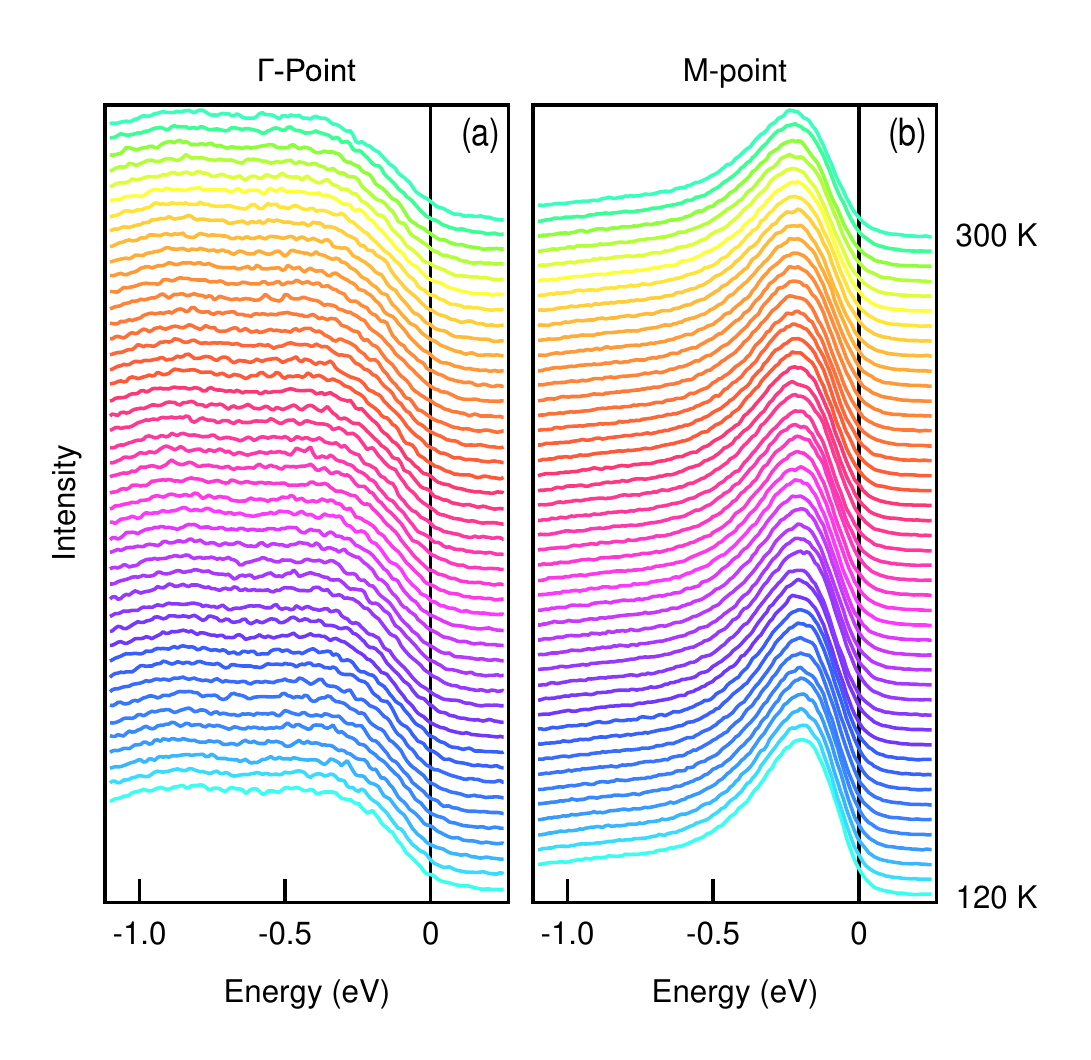}
\caption{\label{fig: Figure5} Temperature-dependent ARPES spectra of Ba-214 from $T=120$~K to $T=300$~K, measured at (a) the  $\Gamma$, and (b) the $M$ points of the BZ.}
\end{figure}

\subsection{Temperature evolution}
We now address the issue, raised in the introduction, of the persistence of the energy gap above $T_N\sim240$~K. Previous theoretical and experimental studies \cite{arita_ab_2012,Hsieh2012,Li2013} have claimed significant Slater-type contributions to the stability of the gap, which should then collapse in the paramagnetic phase. We collected data over a broad temperature range, from well below ($120$~K) to well above ($300$~K) $T_N$. Figure 5 displays temperature-dependent ARPES spectra measured at the $\Gamma$ and $M$ points of the BZ. The leading edge of the spectra exhibits a trivial thermal broadening, but no indications that the gap closes at $T_N$. The energy gap appears to be robust even in the absence of long-range magnetic order. Therefore, the ARPES data do not support a Slater picture, at least in its simplest form.

\section{Electronic structure calculations}
\subsection{First-principles calculations}

\begin{figure}
\centering
\includegraphics[width=86mm]{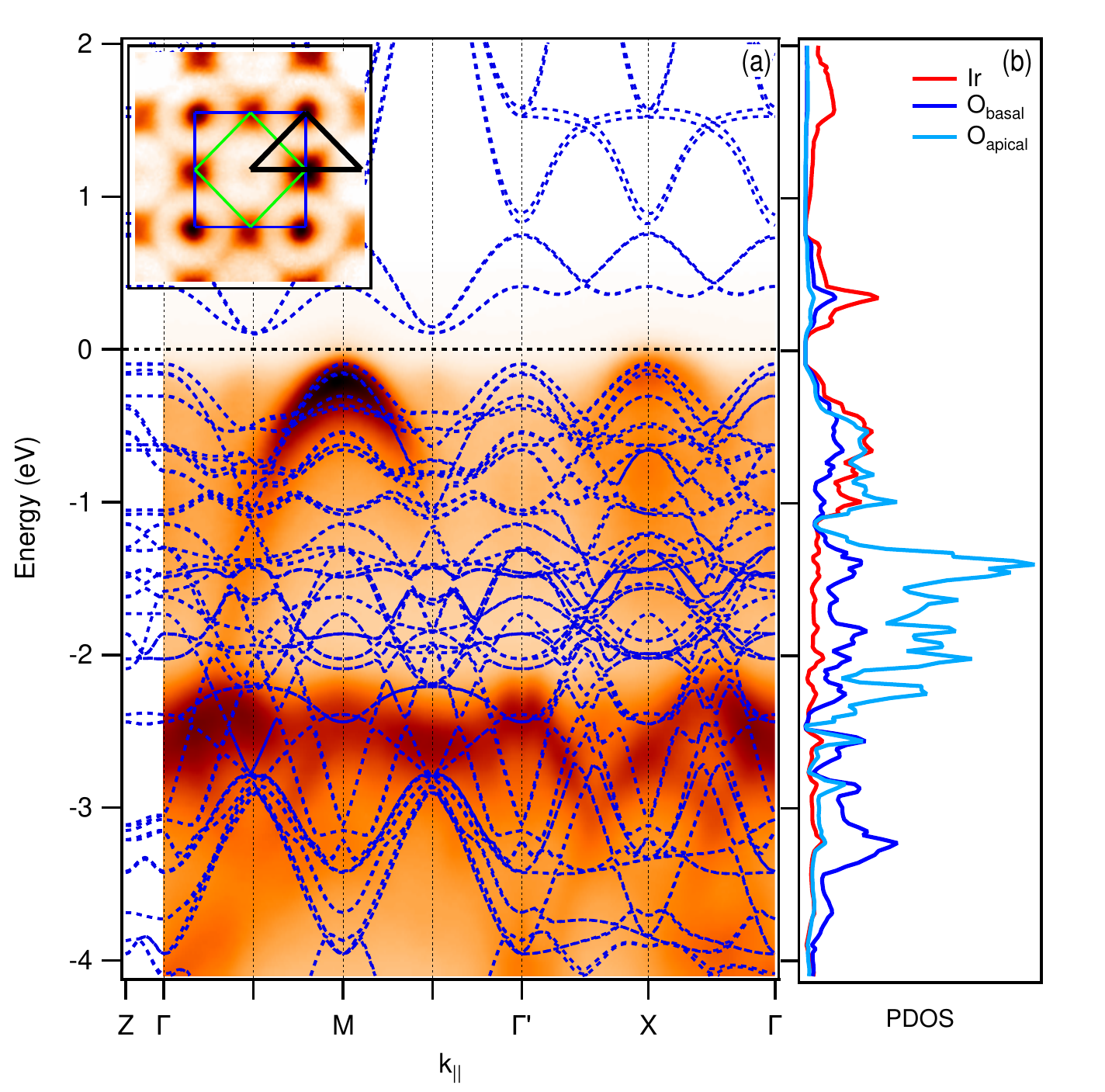}
\caption{\label{fig: Figure6}  (a) The LDA+U+SO band structure is superimposed on an ARPES intensity map, where an integral Shirley background has been subtracted. The bands were calculated for the AFM configuration with $U=3$~eV and $J_H=0.4$~eV. Apical oxygen states between $-1$ and $-2$~eV are not resolved in the experiment. (b) Partial densities of states. Red, dark and light blue lines correspond to Ir $5d$, in-plane oxygen $2p$ and apical oxygen $2p$ states, respectively.}
\end{figure}

We performed an LDA calculation including the on-site Coulomb and spin-orbit interactions (LDA+$U$+SO) \cite{shorikov_magnetic_2005}. For this purpose, the linear muffin-tin orbital approach in the atomic sphere approximation  (Stuttgart LMTO47 code)\cite{andersen_linear_1975} was used, with crystal structure data taken from Ref.~\cite{Okabe2011}. We include Ba($6s,6p,5d$), Ir($6s,6p,5d$) and O($2s,2p$) states in the orbital basis set. Both the ferromagnetic (FM) and AFM configurations were simulated for different sets of on-site Ir $5d$ Coulomb repulsion $U$ and intra-atomic exchange interaction $J_H$. The FM configuration is calculated with one Ir atom per unit cell.
In order to reproduce the AFM order observed in~\BIO, we have used a supercell containing four Ir atoms. The primitive lattice vectors in units of $a=4.03$~{\AA} are $(0,2,0)$, $(2,0,0)$ and $(0.5,0.5,-1.65)$. In the AFM configuration, $U=3$~eV and $J_{H}=0.4$~eV produced the correct energy gap value of 140~meV. The same values were recently used in Ref.~\cite{Comin2012}. 

The calculated band structure is illustrated in Fig.~6(a), superimposed on the ARPES data. Overall, DFT yields many more states than ARPES can individually resolve, which probably explains the absence of sharp quasiparticle features in the spectra. The partial densities of states of Fig.~6(b) show that  Ir $5d$ and $2p$ in-plane oxygen states are strongly hybridized at the top of the valence band, while the bottom of the conduction band is mainly formed by Ir 5$d$ electrons. The $2p$ states of the apical oxygen atoms are confined between $-1$~eV and $-2$~eV, due to a limited overlap with the Ir $5d$ orbitals. These states have a non-negligible dispersion along the $c$-axis. The band observed in ARPES around $-3$~eV corresponds to in-plane $2p$ oxygen states.

We now briefly discuss the implications of the DFT results for the magnetic properties of Ba-214. Firstly, we emphasize that the occupied states close to the gap at the $\Gamma$ and $M$ points strongly depend on the specific magnetic configuration. For instance, in the FM state the top of the valence band is found at $E\simeq-0.7$~eV (not shown). Moreover, a FM insulating ground state with the correct energy gap could only be obtained for unreasonably large values of $U$ and $J_H$. Therefore, the inter-site exchange interaction plays a decisive role in determining the band structure close to the Fermi level. Depending on the strength of the SO coupling, either an $LS$ coupling scheme with angular momentum operators $\vec{S}$ and $\vec{L}$, or a $jj$ coupling scheme with operator $\vec{J}$, are well defined. When the SO coupling and the exchange interaction are comparable, neither the $LS$ nor the $jj$ scheme are valid, and an intermediate coupling theory should be developed \cite{shorikov_magnetic_2005}. Practically, it means that the occupation matrix is neither diagonal in the \{$LS$\} nor in \{$j m_{j}$\} orbital basis. Such a situation is realized in Ba-214, where the SO strength $\lambda \sim 0.48$ eV and J$_{H} \sim 0.4$~eV.

Within an \{$LS$\} eigenstates basis we obtain for the spin and orbital magnetic moments $M_{S}$= 0.12~$\mu_{B}$ and $M_{L}$ = 0.33$~\mu_{B}$, respectively. The resulting total magnetic moment is therefore $M_{LS}=(2M_{S} + M_{L})=0.57~\mu_{B}$, in reasonable agreement with the value $M=0.36~\mu_{B}$ from magnetic susceptibility measurements \cite{Okabe2011}. The total magnetic moment calculated within a \{$j m_{j}$\} basis is equal to $M_{J}=0.43~\mu_{B}$, in better agreement with the experimental value. However, one should note that in both basis sets there are large non-diagonal elements of the occupation matrix that do not contribute to the expectation value of the magnetic moment. Therefore, an intermediate coupling scheme should be used to correctly describe the magnetism of Ba-214. We leave such a consideration for a future investigation.

\subsection{Tight Binding Approach}
\begin{figure}
\centering
\includegraphics[width=90mm]{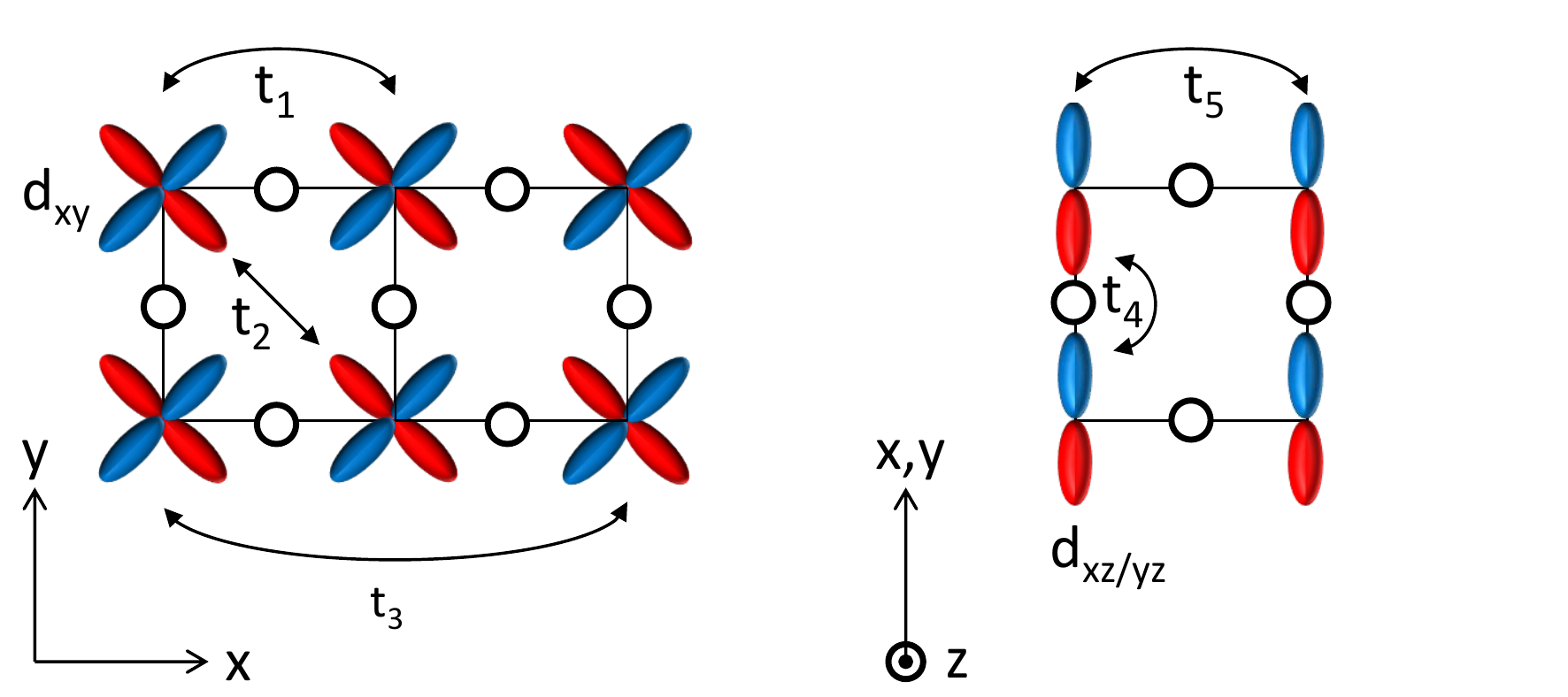}
\caption{\label{fig: Figure7} Effective hopping terms for the Ir $t_{2g}$ states used in the TB model: (a) $d_{xy}$ orbitals; (b) $d_{xz/yz}$ orbitals. Empty circles represent oxygen ions.}
\end{figure}
In order to gain more direct insight in the interplay of orbital ordering, SOC and correlation effects, we also performed a model tight-binding (TB) calculation, along the lines of Refs.~{\bf} \cite{Watanabe2010} and {\bf} \cite{Martins2011}. We included the whole Ir 5$d$ $t_{2g}$ and $e_g$ manifolds, and a set of effective hopping terms describing the hybridization between the Ir $5d$ and O $2p$ electrons, as illustrated in Fig. \ref{fig: Figure7}. The tight binding Hamiltonian is:
\begin{equation}\label{equ: hamiltonian}
H=H_{0} + H_{SO}~.
\end{equation}
\noindent $H_{0}$ includes the kinetic term $T=\Sigma_{k,\nu}~\varepsilon_{\nu}(k)c^{\dagger}_{k\nu}c_{k\nu}$, where $c^\dagger_k$ and $c_k$ are the fermion creation and annihilation operators and $\nu$ spans  the Ir $5d$ manifold, and an octahedral CEF parametrized by $10Dq=E(e_g)-E(t_{2g})$. Since we are primarily interested in the occupied states for a comparison with ARPES, we focus on the $t_{2g}$ levels. The relevant hopping terms are schematically illustrated in Fig.~7.  The form of the $\varepsilon_{\nu}$'s is dictated by the symmetry of the system \cite{Watanabe2010}:
\begin{eqnarray}\label{equ: kinetic terms}
\varepsilon_{xy}&=&-2t_{1}(cosk_x+cosk_y)-2t_2cosk_xcosk_y\nonumber\\
&&-2t_{3}(cos2k_x+cos2k_y)~;\nonumber\\
\varepsilon_{xz}&=&-2t_{4}cosk_x-2 t_{5}cosk_y~;\nonumber\\
\varepsilon_{yz}&=&-2t_{5}cosk_x-2 t_{4}cosk_y~.
\end{eqnarray}
\noindent The SO coupling term for the $5d$ orbitals is: $H_{SO}=\lambda_{5d}\ \vec{L} \cdot \vec{S}$.
We introduce electron correlations in the model in a phenomenological way, by imposing AFM order. This is achieved by an additional Zeeman term with an in-plane staggered magnetic field:
\begin{equation}\label{equ: zeeman term}
H_{AFM}=B\sum_{i,\nu} e^{i\vec{Q}\cdot\vec{r_i}}(c^{\dag}_{i\nu \uparrow}c_{i\nu \downarrow}+c^{\dag}_{i\nu \downarrow}c_{i\nu \uparrow})~,
\end{equation}
where $\vec{Q}=(\pi,\pi)$ is the AFM ordering vector, and the sum is over the Ir sites $i$ and the three $t_{2g}$ orbitals. The effect of~$H_{AF}$ is to fold all bands into the smaller $c(2\times2)$ BZ, and to open gaps at the AFM zone boundaries.

The band structures produced by the various terms of the hamiltonian are plotted in Fig.~8(a-c), along the same $\Gamma X'M'\Gamma$ contour of Fig.~4(a,b). The parameters of the model are summmarized in Table II. Panels
8(a'-c') schematically illustrate the local electronic structure at the Ir site. Figure 8(a) shows the band dispersion in the presence of the octahedral CEF. The empty $e_g$ and the partially filled $t_{2g}$ manifolds are well separated by $10Dq$, and the system is metallic. Adding the SO interaction, in panels 8(b,b'), mixes the CEF states. The $t_{2g}$ states are split into a 4-fold degenerate, fully occupied $j_{eff}=3/2$ (blue), and a doubly degenerate, half-filled, $j_{eff}=1/2$ manifold (red), but the system remains metallic. Panels 8(c,c') illustrate the further band splitting induced by $H_{AFM}$, namely of the $j_{eff}=1/2$ band into $m_j=-1/2$ and $m_j=1/2$ subbands, which effectively simulates the opening of a correlation gap between an occupied lower Hubbard band and an empty upper Hubbard band.

\begin{table}
\caption{\label{tab: Table2} The set of parameters (in eV) of the TB model used for the calculated band structure of Fig.~8 and 9.}
\centering
\begin{tabular}{c|c|c|c|c|c|c}
\br
  $\lambda_{5d}$ & $t_{1}$ & $t_{2}$ & $t_{3}$ & $t_{4}$ & $t_{5}$ & $B$ \\
  \hline
  0.7& 0.5 & 0.1 & 0.03 & 0.27 & 0.01 & 0.1 \\
\br  
\end{tabular}
\end{table}

\begin{figure}
\includegraphics[width=76mm]{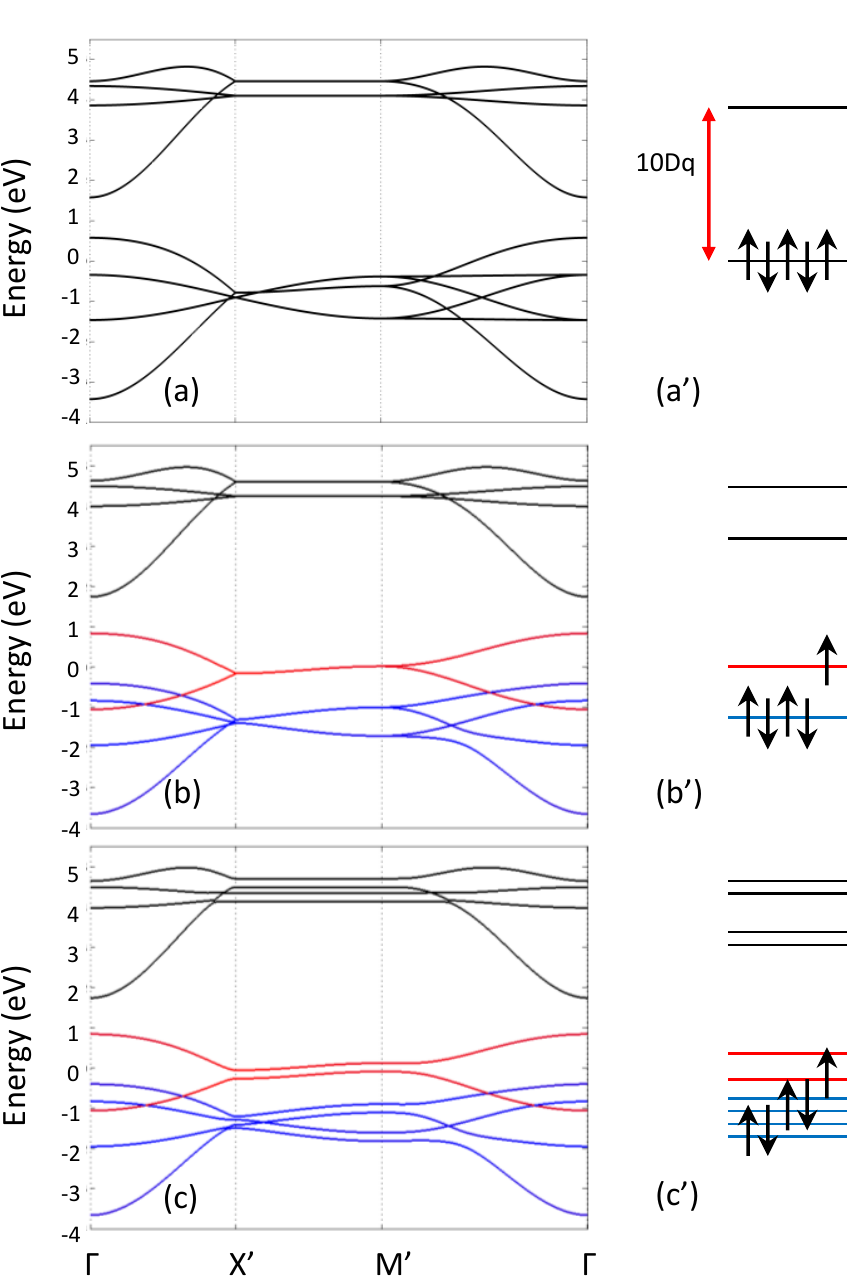}
\centering
\caption{\label{fig: Figure8} The calculated TB band structure is shown in (a) in the presence of an octahedral CEF. In (b) the addition of SO coupling rearranges the bands into $j_{eff}=3/2$ (blue) and $j_{eff}=1/2$ (red) states.
A staggered magnetic field splits all states in (c), namely within the half-filled $j_{eff}=1/2$ band, simulating the opening of the Mott gap. Panels (a'-c') are corresponding schematic pictures of the local electronic structure at the Ir sites.
}
\end{figure}

\begin{figure}
\centering
\includegraphics[width=84mm]{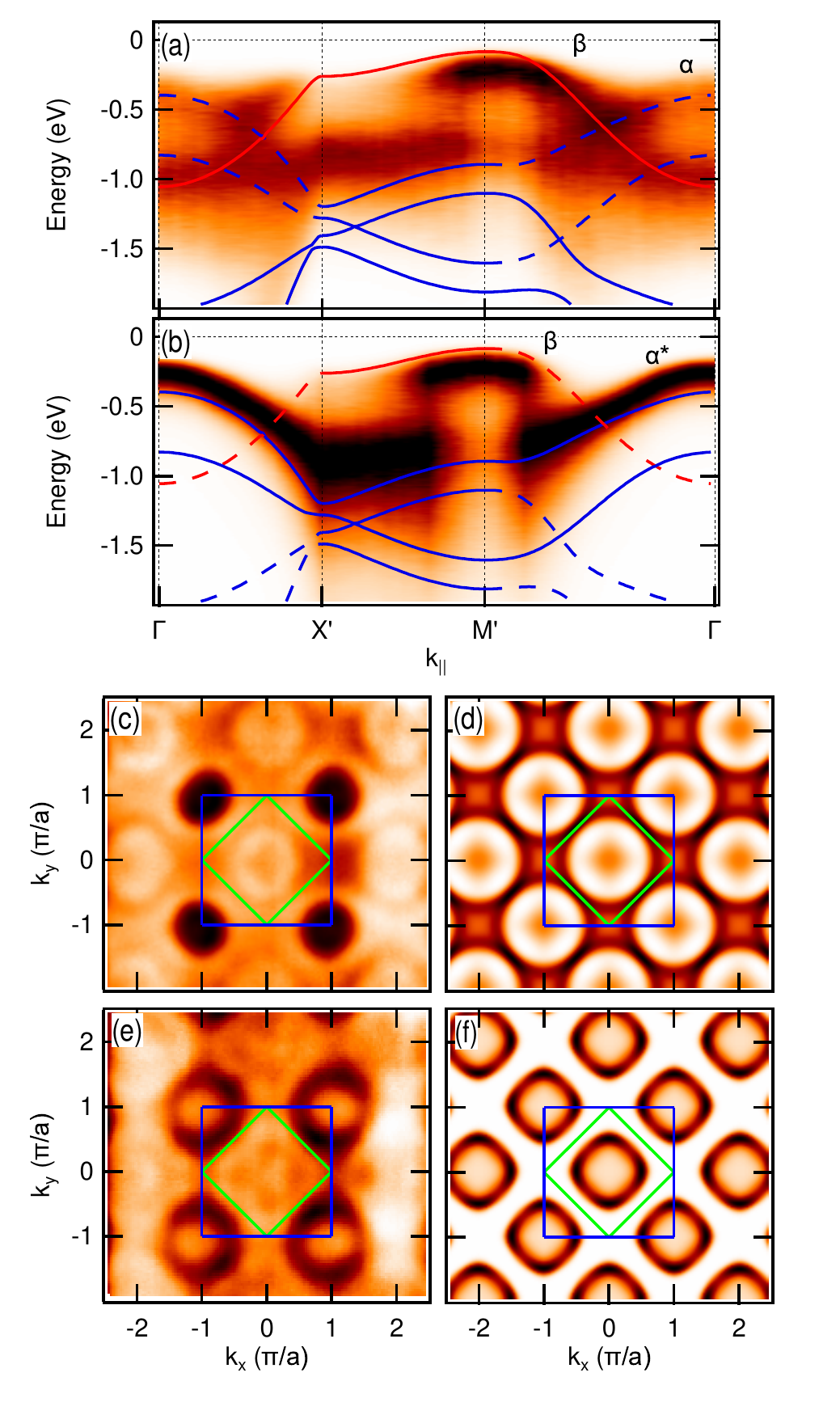}
\caption{\label The calculated TB band structure is superimposed on the experimental ARPES data, along two equivalent contours in the 1st (a) and in an adjacent (b) $c(2\times2)$ BZ, as in Fig.~4(b) and (c).  Folded bands are indicated by dashed lines. The color code is the same as in Fig.~8.  Panels (c,d) show the experimental (c) and calculated (d) CE maps for $E=-0.35$~eV.  The corresponding CE maps for $E=-0.7$~eV are shown in (e,f).}
\end{figure}

The calculated TB band structure is compared with the ARPES data in Fig.~9 for the set of parameters of Table \ref{tab: Table2}. Although the TB parameters, namely the external magnetic field, should not be taken too literally, they do provide a useful description of the electronic structure. Panel 9(a) and 9(b) show data along equivalent contours in the 1st and in the adjacent $c(2\times2)$ BZs, as in Fig.~4(b) and (c). Folded bands are indicated by dashed lines. The good agreement with the data
substantiates the description of the bands given in Section III, namely the assignment of the top of the valence band to states of $j_{eff}=1/2$ character.
The good agreement of the TB model with the experiment is confirmed by a comparison of the experimental and calculated CE maps shown in Fig.~9 (c,d) and Fig.~9 (e,f), for $E=-0.35$~eV (c,d) and $E=-0.7$~eV (e,f). The experimental $\alpha$ ($\alpha^*$) and $\beta$ features are well reproduced. Of course, the TB model does not yield any information on the spectral weight, and therefore all $c(2\times2)$ BZs are equivalent.

\section{Summary}

We have measured the electronic structure of the perovskite iridate~\BIO~(Ba-214) by ARPES on high-quality single crystal samples grown under high pressure. A comparison of spectra measured at non-equivalent locations of reciprocal space allows us to unambiguously identify the $j_{eff}=3/2$ and $j_{eff}=1/2$ subbands into which the Ir $5d~t_{2g}$ manifold is split by the SO interaction. The experimental data are well reproduced by an LDA + $U$ + SO calculation for an AFM configuration. A satisfactory agreement is also achieved by a simple empirical tight-binding model. The overall band dispersion is similar to that of the sister compound Sr-214. The electronic structure is therefore rather insensitive to the rotational distortion of the IrO$_6$ octahedra, which is present in Sr-214 but not in Ba-214. This observation contrasts with the behavior of $3d$ TM perovskites, such as the rare earth nickelates RNiO$_3$, where the tilting of the octahedra, which affects the orbital overlap, strongly influences the band dispersion, as well as transport and magnetic properties \cite{Torrance}. This lends support to a proposed scenario for the iridates, where the effective hopping parameters are less sensitive to the distortion due to the strong mixing of $d_{xy}$, $d_{xz}$ and $d_{yz}$ orbital characters induced by the SO interaction \cite{Katukuri2012}. 

We have also found that the bands measured by ARPES are folded, with reduced intensity, into a smaller $c(2\times 2)$ BZ, producing a characteristic checkerboard intensity distribution. Band folding has also been observed in Sr-214 \cite{Kim2008}, and attributed to the effect of the structural distortion. Very sensitive x-ray diffraction measurements with synchrotron radiation rule out a structural distortion in the bulk of Ba-214 \cite{McMorrow}. Our results are consistent with the periodicity of the AFM structure, and could indicate scattering of the quasiparticles by the corresponding superlattice potential. However, we cannot exclude that a structural distortion develops at the surface, which has indeed been suggested by low-energy electron diffraction (LEED) measurements on thin film samples \cite{SHENprivatecomm}. Accurate surface x-ray diffraction experiments are necessary to determine the surface structure of Ba-214, and resolve this issue.

Finally, we have studied the evolution with temperature of the energy gap, and found no notable variations, namely around the magnetic ordering temperature $T_N=230~K$. The gap remains open well into the paramagnetic phase. The ARPES data are therefore more consistent with a Mott than with a Slater scenario, even if our calculations indicate a clear influence of magnetic order on the size of the gap.

\section{Acknowledgments}
We gratefully acknowledge insightful discussions with V.I. Anisimov, A.O. Shorikov, B. J. Kim, D.F. McMorrow, S. Boseggia and C. Tournier-Colletta. Special thanks are due to K. M. Shen for sharing with us his unpublished data. The work at Lausanne is supported by the Swiss NSF. L.M. is supported by the Swiss NSF Grant N PA00P21-36420. The work of V.V.M. is supported by the grant program of President of Russian Federation MK-5565.2013.2 and the contract of the Ministry of education and science of Russia N 14.A18.21.0076. The Advanced Light Source is supported by the Director, Office of Science, Office of Basic Energy Sciences, of the U.S. Department of Energy under Contract No. DE-AC02-05CH11231.



\end{document}